\RequirePackage[2020-02-02]{latexrelease}
\documentclass[aps,showpacs,superscriptadress,preprint]{revtex4}
\usepackage{graphicx}
\usepackage{amsmath}
\begin{document}

\title{Antikaon condensation in magnetized neutron star matter within the framework of the $\sigma$-cut scheme}

\author{
Fei Wu$^1$ \footnote{f.eiwu@outlook.com}  }
 \affiliation{
\small 1. Xingzhi College, Zhejiang Normal University, Jinhua, 321004, Zhejiang, China}

\author{
Chen Wu$^1$\footnote{wuchenoffd@gmail.com} }
 \affiliation{
\small 1. Xingzhi College, Zhejiang Normal University, Jinhua, 321004, Zhejiang, China}

\begin{abstract}
This study investigates the effects of strong magnetic fields on antikaon condensation in neutron star matter using the extended FSUGold model model.
It is found that the presence of strong magnetic fields alters the threshold density of antikaon condensation significantly, which means the threshold density  of antikaon condensation is shifted to higher density compared with the magnetic field-free case.
 In the presence of strong magnetic fields, the equation of state (EoS) becomes stiffer than that of the zero field case.  The effects of the $\sigma$-cut scheme on  the EoS  are also researched when the appearance of antikaon condensation is occurred.
Through careful choice of the parameter of the $\sigma$-cut scheme, we are able to produce a maximum mass neutron star heavier than 2$M_{sun}$.

\end{abstract}
\maketitle

\section{Introduction}
 Born from the remnants of a supernova explosion, a neutron star exhibits a
range of densities inside its structure, possibly under strong magnetic field \cite{Lattimer:Intro}. The state of matter inside neutron stars is an
unsolved mystery of modern nuclear physics.  The observation of many new  neutron stars like PSR J1614 - 2230 ($M = (1.928\pm 0.017)M_{Sun}$) \cite{demorest2010two},  PSR J0348
  + 0432 ($M = (2.01 \pm 0.04)M_{Sun}$) \cite{antoniadis2013massive} and MSP J0740+6620 ($M = (2.08 \pm 0.07)M_{Sun}$)  \cite{Fonseca_2021, cromartie2020relativistic} not only questions the occurrence of exotic  degrees of freedom but also places strong constraints on the EoS of nuclear matter. These astrophysical observations indicate that the possible lower limit of neutron stars maximum mass
is above $2M_{Sun}$. The structure and composition of a neutron star is determined by the EoS of the strongly interacting constituents.
 It has been gradually realized that particles beyond neutrons,
protons and leptons should also be investigated in the study of neutron stars. In fact, the inner cores of the neutron stars are sources of speculation.
 At high density, kaon condensation, quark deconfinement,  hyperons and Delta resonance are possible to appear and much investigation has been paid to these fields (see \cite{yin2010slowly,van2012fermi,PhysRevD.109.063008,wu2017neutron, ma2022kaon, ma2023kaon, PhysRevC.77.045804, PhysRevC.107.035807}).
  Many complicated  EoSs with these possible exotic
  degrees of freedom  are able to produce a maximum mass neutron star heavier than 2 $M_{sun}$ \cite{PhysRevD.109.063008, ma2022kaon, ma2023kaon, PhysRevC.107.035807}.

  Kaplan and Nelson have suggested that the ground state of hadronic matter might form a negatively charged Kaon Bose-Einstein condensation above a certain critical density \cite{Kaplan:1986yq,Nelson:1987dg}.  When the vacuum mass of the kaon meson is exceeded by the electronic chemical potential,  as the density increases, negatively charged  kaon mesons begin to appear, which helps to maintain charge neutrality.  The effective mass of the $K^{-}$ meson is decreased due to the interaction with the nucleon.
  The occurrence of onset of s-wave $K^-$ condensation is identified by equating the chemical potential of $K^-$ ($\omega_{K^-}$ ) and electron chemical potential $\mu_e$. To maintain the charge neutrality condition, $K^-$ condensates replace the electrons.
  The threshold density for the onset of $K^-$ is highly sensitive to its optical potential. The presence of $K^-$ condensation in neutron star matter has been extensively studied in past literature \cite{ma2022kaon, PhysRevC.60.025803, PhysRevC.63.035802, PRAKASH19971, PhysRevC.53.1416, PhysRevC.64.055805, Malik2021NewEO, PhysRevD.102.123007}. In general, the presence of antikaon condensation tends to soften the EoS at high density and lower the maximum mass of neutron stars.

 On the other hand, the observation of soft gamma repeaters  and  anomalous X-ray pulsars  indicate that the magnitude of magnetic field at the surface of neutron stars is of the order
   $10^{14}$ - $10^{15}$ G \cite{duncan1992formation, thompson1995soft, usov1992millisecond, 1538-4357-486-2-L129, 1538-4357-519-2-L139, kouveliotou1998x}.
    It is far above the critical field strength $B_c^e=4.414 \times 10^{13}$ G for quantization of electrons \cite{broderick2000equation}.
    So far, there is no direct observational evidence for the internal magnetic fields of the star, while it may reach $10^{18}$ G, as estimated in some theoretical works. Recently,  one international cooperative team reported that a binary neutron star merger will leave behind a massive neutron star with a strong magnetic field \cite{xue2019magnetar}.   They found a X-ray transient, which is most likely produced by a binary neutron star merger. The light curve is very likely to be powered by a magnetar.
     So understanding the impact of strong magnetic field on the properties of neutron stars, adopting different approaches,
is a hot topic of research \cite{canuto1968thermodynamic, canuto1968quantum, canuto1968magnetic, canuto1977quantizing, PhysRevC.77.045804, PhysRevC.107.035807}. It is also
interesting to investigate the influence of strong magnetic field on antikaon condensation.

Previous studies have been conducted on neutron star matter containing  the condensation of negatively charged $K^-$ under the influence of strong magnetic
field  \cite{2022IJMPE..3150050K,PhysRevC.77.045804, PhysRevC.107.035807, Dey_2002}. It was found that the threshold of $K^-$ condensation
shifts to higher density in the presence of strong magnetic field and the EoS becomes stiffer. These qualitative features are expected to persist in other models.
In this article, we study the effects of strong magnetic field on the condensation of negatively charged $K^-$ in $\beta$ equilibrium matter  within the framework of the $\sigma$-cut scheme.
To get the EoS of neutron star matter, the relativistic mean field theory is usually applied \cite{Glendenning:1991ic,Ryu:2010zzb,Lopes:2013cpa,glendenning1991reconciliation,broderick2000equation}, which  has achieved great success in the study of the properties of nuclei and nuclear matter.
There are many theoretical models within the relativistic mean field framework. Recently, a plenty of relativistic mean field models were researched to test the range of resulting neutron star masses, where 14 of them could result in masses within the range (1.93 - 2.05)$M_{Sun}$ , of which only two of the models could satisfy the mass constraint  \cite{Dutra:2015hxa}.
In this context, we are going to utilize the FSUGold model \cite{todd2005neutron} as an example.
It has been successfully applied to reproduce properties of finite  spherical nuclei and nuclear matter \cite{piekarewicz2007validating}.
 Nevertheless, the maximum mass of the neutron stars obtained by  the FSUGold model is too small, well below the maximum observed mass (2.01 $M_{sun}$) \cite{Antoniadis:2013pzd}.
 The problem is that the EoS generated by this model is not stiff enough.  Some researchers argued that this result rules out the FSUGold model as a suitable description of  high density nuclear matter.
  However, some researchers have proposed a $\sigma$-cut scheme \cite{Maslov:2015lma} to modify the $\sigma$ meson self-interaction term at high density, and is able to quench the  obvious decreasing of the effective mass of nucleons at high density.  This procedure offers a simple way to stiffen the EOS at high density.  However, the effect of the condensation of antikaon  is not included in the study \cite{Maslov:2015lma}.
  It is well known that the introduction of antikaon condensation leads to a softer EoS, thus producing a neutron star with a smaller mass.
  It leaves  the open question of whether the $\sigma$-cut scheme is still effective when antikaon condensation under strong magnetic field is considered. So we will utilize the $\sigma$-cut scheme as a tentative approach in this study.

This paper is organized as follows. First, we briefly describe the FSUGold model for neutron star matter with antikaon
condensation in the presence of strong magnetic field. Then  we show and discuss the numerical results in this model and make a systematic investigation   within the framework of the $\sigma$-cut scheme.  Finally, some conclusions are provided.

\section{Theoretical framework}
The starting point of the extended FSUGold model is the Lagrangian density \cite{todd2005neutron,wu2011strange,rabhi2009quark}
 \begin{align}\label{Lagrangian}
\mathcal{L} = &\sum_{B=p,n}
\bar{\psi}_B[i\gamma^{\mu}\partial_\mu-q_B\gamma^{\mu}A_{\mu}-m_{B}+g_{\sigma
B}\sigma-g_{\omega B}\gamma^\mu\omega_\mu \nonumber \\ &
- g_{\rho B}
\gamma^\mu \vec{\tau}_B \cdot \vec{\rho}^\mu-\frac{1}{2}\mu_N\kappa_B\sigma_{\mu\nu}F^{\mu\nu}
]\psi_B  \nonumber \\ &
+\frac{1}{2}\partial_{\mu}\sigma\partial^{\mu}\sigma -\frac{1}{2}m_\sigma^2 \sigma^2 -
\frac{\kappa}{3!}(g_{\sigma N} \sigma)^3 -\frac{\lambda}{4!}(g_{\sigma N} \sigma)^4 \nonumber \\ &
-\frac{1}{4}\Omega_{\mu\nu}\Omega^{\mu\nu}+\frac{1}{2}m_{\omega}^2\omega_\mu\omega^\mu+
\frac{\zeta}{4!}(g_{\omega N}^2 \omega_\mu \omega^\mu)^2  \nonumber\\ &
-\frac{1}{4}\vec{G_{\mu\nu}}\vec{G^{\mu\nu}}+\frac{1}{2}m_{\rho}^2\vec{\rho_\mu} \cdot \vec{\rho^\mu}
 + \Lambda_v (g_{\rho N}^2 \vec\rho_\mu \cdot \vec\rho^\mu )(g_{\omega N}^2 \omega_\mu \omega^\mu)  \nonumber  \\ &
-\frac{1}{4}F_{\mu\nu}F^{\mu\nu}+ \sum_{l=e, \mu} \bar{\psi}_l[i\gamma^{\mu}\partial_\mu-m_{l}-q_l\gamma^{\mu}A_{\mu} \nonumber ]\psi_l  \\ &
 + D_\mu^{*}{K^*}D^\mu K - m^{*2}_K K^*K,
\end{align}
where $\psi_B$ and $\psi_l$ represent the nucleons and leptons, respectively. $\vec{\tau}_B$ is the isospin operator for the ${\vec{\rho}}^\mu$ meson fields.
The interaction of anomalous magnetic moments of baryons with magnetic fields is given by the last term under the first summation in Eq.(\ref{Lagrangian}).
Here, $F^{\mu\nu}$ is the electromagnetic field tensor, $\sigma_{\mu\nu}=[\gamma_\mu, \gamma_\nu]/2$, $\mu_N$ is  the nuclear magneton. The anomalous magnetic moments of nucleons are given by $\kappa_p = 1.7928$ and $\kappa_n = -1.9130$. The covariant derivatives in Eq.(\ref{Lagrangian}) are given by
 $ D_{\mu}=  \partial_\mu + ig_{\omega K}\omega_\mu + ig_{\rho K} \vec\tau_K \vec\rho_\mu + iq_K A_{\mu}$. The effective masses  $ m_K^* $ of antikaons  are given by     $ m_K^* = m_K - g_{\sigma K}\sigma $.  There is no interaction term involving magnetic moments in the kaonic Lagrangian density because kaons having zero spin angular momentum do not possess magnetic moments. The electric charges of particles are $q_e = q_\mu = q_{K^-} =-e$, $q_n = 0$ and $q_p = e$.
$m_B$,  $m_l$ and $m_K$ stand for the masses of nucleons, leptons and kaons, respectively. $\sigma$, $\omega_\mu$ and $\vec{\rho}_\mu$ are the meson fields with masses $m_\sigma$, $m_\omega$  and $m_\rho$, respectively. $\Lambda_v$ is introduced to modify the density dependence of symmetry energy.
Under the effect of this magnetic field, the motion of the charged particles is Landau quantized in the plane perpendicular to the direction of field, the momentum in the perpendicular direction being $p_\perp = 2\nu q B$, where $\nu$ is the Landau level.

The gauge mesonic contributions in Eq.(\ref{Lagrangian}) contain the field strength tensors:
\begin{align}\label{4}
\begin{split}
    \Omega_{\mu\nu}= \partial_\nu\omega_\mu - \partial_\mu\omega_\nu,\\
    \vec{G}_{\mu\nu}= \partial_\nu \vec{\rho}_\mu - \partial_\mu\vec{\rho}_\nu.
\end{split}
\end{align}

In the relativistic mean field approximation, the three meson field equations in the presence of antikaon condensate and magnetic field are
\begin{equation}
\begin{aligned}
&m_{\sigma}^{2}\sigma+\frac{1}{2}{\kappa}g_{\sigma N}^{3}{\sigma}^{2}+\frac{1}{6}{\lambda}g_{\sigma N}^{4}{\sigma}^{3}= g_{\sigma B} ({\rho}_{p}^{S} + {\rho}_{n}^{S})
+g_{\sigma K} \frac{m_K^*}{\sqrt{{m_K^*}^2+|q_{K^-}|B}} {\rho}_{K^-},\\
&m_{\omega}^{2}{\omega}+\frac{\zeta}{6}g_{\omega N}^{4}{\omega}^{3}+2{\Lambda_{\nu}}g_{\rho N}^{2}g_{\omega N}^{2}{\rho}^{2}{\omega} = g_{\omega B}({\rho}_{p} + {\rho}_{n})- g_{\omega K}{\rho}_{K^-},\\
&m_{\rho}^{2}{\rho}+2{\Lambda}_{\nu}g_{\rho N}^{2}g_{\omega N}^{2}{\omega}^{2}{\rho}={g_{\rho B}}({\rho}_{p} - {\rho}_{n})-{\frac{g_{\rho K}}{2}{\rho}_{K^-}},
\end{aligned}
\end{equation}
where $\rho_{p}$($\rho_{n}$) and $\rho_{p}^{S}$($\rho_{n}^{S}$) are the proton(neutron) density and the scalar density, respectively and the Kaon density $\rho_{K^-} = 2\sqrt{{m_K^*}^2 + |q_{k^-}|B}K^{*}K$.

The main effect of the magnetic field is Landau quantization. The energy spectra for neutrons, protons and leptons are given by  \cite{rabhi2009quark,broderick2002effects}
 \begin{align}
E^n_{s} =& \sqrt{{k_z}^2+ (\sqrt{{m_B^*}^2+k_x^2+k_y^2}-s\mu_N\kappa_n B)^2} \nonumber + g_{\omega B}\omega - g_{\rho B} \rho, \nonumber \\
E^p_{\nu,s} =& \sqrt{{k_z}^2+ (\sqrt{{m_B^*}^2+2\nu|q_p|B}-s\mu_N\kappa_p B)^2} \nonumber + g_{\omega B}\omega + g_{\rho B} \rho, \nonumber \\
E^l_{\nu,s} =& \sqrt{{k_z}^2+ {{m_l}^2+2\nu|q_l|B}},
 \end{align}
where $\nu=0,1,2,3$ ... enumerates the Landau levels of  fermion i with electric charge $q_i$
( $i = p, e, or\mu$) ;  The quantum number $s$ is +1 for spin-up and
-1 for spin-down cases.
 When $\kappa_B$ is set to zero, the effect of the AMMs is switched off.
The scalar density, baryon number density and the kinetic energy density of neutrons  are given by:
\begin{equation}
\begin{aligned}
       \rho^s_n =& \frac{M_N^*}{4\pi^2} \sum_{s = \pm 1} \left[k_{f,s}^n E_f^n - {( M_N^* - s\mu_N \kappa_n B)}^2 \ln{\left(\frac{k_{f,s}^n + E_f^n}{ M_N^* - s\mu_N \kappa_n B}\right)}\right] \label{8} \\
\end{aligned}
\end{equation}
\begin{equation}
\begin{aligned}
\rho_{n} =\frac{1}{2\pi ^{2}}\sum_{s=\pm 1}\left\{ \frac{1}{3}k_{f,s}^{n3}-
\frac{s\mu_B \kappa _{n}B}{2}\left[ \left( M_{N}^{\ast}-s{\mu_B\kappa _{n}B}\right)
k_{f,s}^{n}\right. \right. \left. \left. + E_{f}^{n2}\left( \arcsin \frac{M_{N}^{\ast}-s{\mu_B\kappa _{n}B}
}{E_{f}^{n}}-\frac{\pi}{2}\right) \right] \right\} ,  \label{eq:rhovn}
\end{aligned}
\end{equation}
\begin{equation}
\begin{aligned}
\varepsilon _{n}  =&\frac{1}{4\pi ^{2}}\sum_{s= \pm 1}\left\{ \frac{1}{2}%
k_{f,s}^{n}E_{f}^{n3}-\frac{2s\mu_N\kappa _{n}B}{3}E_{f}^{n3}\left( \arcsin \frac{
M_{N}^{\ast}-s{\mu_N\kappa _{n}B}}{E_{f}^{n}}-\frac{\pi}{2}\right) \right.  \\  & -\left(
\frac{s{\mu_N\kappa _{n}B}}{3}+\frac{M_{N}^{\ast}-s{\mu_N\kappa _{n}B}}{4}\right)
\\  &
\times  \left. \left[ \left( M_{N}^{\ast}-s{\mu_N\kappa _{n}B}\right)
k_{f,s}^{n}E_{f}^{n}+\left( M_{N}^{\ast}-s{\mu_N\kappa _{n}B}\right) ^{3}\ln
\left\vert \frac{k_{f,s}^{n}+E_{f}^{n}}{M_{N}^{\ast}-s{\mu_N\kappa _{n}B}}
\right\vert \right] \right\} , \\
\end{aligned}
\end{equation}
where $k_{f,s}^n$ represents the Fermi momentum of neutrons, which is related to the Fermi energies $ E_f^n $ as ${E_{f}^{n}}^{2} = k_{f,s}^{n2}+( M_{N}^* -s\mu_N\kappa _{n}B)^{2}$,
and $ M_{N}^* $ is the effective mass of nucleons, which is defined as $ M_{N}^* = M_{N} - g_{\sigma N}\sigma $.

Similarly for protons, those expressions are given by
\begin{equation}
\rho_{p}^{s} = \frac{{q_{p}B}M_{N}^{\ast}}{2\pi ^{2}}\sum_{\nu}\sum_{s= \pm 1}
\frac{\sqrt{M_{N}^{\ast 2}+2\nu {q_{p}B}}-s\mu_N\kappa _{p}B}{\sqrt{M_{N}^{\ast
2}+2\nu {q_{p}B}}}\ln \left\vert \frac{k_{f,\nu ,s}^{p}+E_{f}^{p}}{\sqrt{
M_{N}^{\ast 2}+2\nu {q_{p}B}}-s\mu_N\kappa _{p}B}\right\vert ,
\end{equation}
\begin{equation}
\begin{aligned}
\rho_{p} = \frac{{q_{p}}B}{2\pi ^{2}}\sum_{\nu}\sum_{s = \pm 1}k_{f,\nu
,s}^{p},
\end{aligned}
\end{equation}
\begin{equation}
\begin{aligned}
\varepsilon _{p}  =\frac{{q_{p}B}}{4\pi ^{2}}\sum_{\nu}\sum_{s= \pm 1}\left[
k_{f,\nu ,s}^{p}E_{f}^{p}+\left( \sqrt{M_{N}^{\ast 2}+2\nu {q_{p}B}}-s\mu_N\kappa
_{p}B\right) ^{2}\ln \left\vert \frac{k_{f,\nu ,s}^{p}+E_{f}^{p}}{\sqrt{%
M_{N}^{\ast 2}+2\nu {q_{p}B}}-s\mu_N\kappa _{p}B}\right\vert \right].
\end{aligned}
\end{equation}
The fermi momentum, $k_{f,\nu,s}^p$ for the protons with spin $s$ is related to the Fermi energy, $E_f^p$ through the relation
${E_{f}^{p}}^{2} =k_{f,\nu ,s}^{p2}+( \sqrt{M_{N}^{\ast 2}+2\nu {q_{p}B }}-s\kappa _{p}B) ^{2}$.

The expressions for the number density and the kinetic energy density of leptons have the same form as that of protons but leptons are noninteracting and anomalous magnetic moment of leptons
is not considered here:
\begin{equation}
\rho_{l}=\frac{\left\vert {q_{l}}\right\vert B}{2\pi^{2}}\sum_{\nu}
\sum_{s = \pm 1}k_{f,\nu ,s}^{l},
\end{equation}

\begin{equation}
   \varepsilon _{l} =\sum_{l=e,\mu}\sum_{\nu}\sum_{s = \pm 1}\frac{\left\vert {q_{l}%
}\right\vert B}{4\pi ^{2}}\left[ k_{f,\nu ,s}^{l}E_{f}^{l}+\left( m_{l}^{2}{+
}2\nu \left\vert {q_{l}}\right\vert {B}\right) \ln \left\vert \frac{k_{f,\nu
,s}^{l}+E_{f}^{l}}{\sqrt{m_{l}^{2}{+}2\nu \left\vert {q_{l}}\right\vert {B}}}
\right\vert \right],
\end{equation}
where  $k_{f,\nu,s}^l$ is the Fermi momenta of leptons, which is related to the Fermi energies $E_f^l$ as ${E_{f}^{l}}^{2} = k_{f, \nu, s}^{l2}+ m_{l}^{2} + 2\nu |q_l| B$.

The chemical potentials of nucleons and leptons are given by
\begin{eqnarray}
\mu_{p} &=&E_{f}^{p}+g_{\omega N}\omega +g_{\rho N}{\rho_{}},
\label{eq:mup} \\
\mu_{n} &=&E_{f}^{n}+g_{\omega N}\omega -g_{\rho N}{\rho_{}},
\label{eq:mun} \\
\mu_{l} &=&E_{f}^{l}.
\end{eqnarray}%

The chemical potential of s-wave condensates of (anti)kaons is given by:
\begin{align}\label{kaon chemical}
    \mu_{K^-} =& \; \sqrt{m_K^{*^2}+ |q_{K^-}|B} - g_{\omega K}\omega - \frac{1}{2}g_{\rho K}\rho.
\end{align}

For the neutron matter with baryons,  charged leptons and antikaons, the $\beta$-equilibrium conditions are guaranteed with the following relations of chemical potentials for different particles:
\begin{align}
    \mu_{K^-} =& \; \mu_e = \; \mu_\mu = \; \mu_n- \mu_p,
\end{align}
and the charge neutrality condition is fulfilled by
\begin{equation}
   \rho_p - \rho_e - \rho_\mu - \rho_{K^-} = 0.
\end{equation}

 We solve the equations listed above self-consistently at a given baryon density in the presence of antikaon condensation and strong magnetic fields. The total energy density of neutron star matter is given by
\begin{equation}
\begin{aligned}
   \varepsilon =&  \varepsilon_p + \varepsilon_n + \varepsilon_l +\frac{1}{2}m_{\omega}^{2}\omega^{2}  +\frac{\zeta}{8}g_{\omega N}^{4}\omega^{4}+\frac{1}{2}m_{\sigma}^{2}\sigma^{2}+\frac{\kappa}{6}g_{\sigma N}^{3}\sigma^{3}+\frac{\lambda}{24}g_{\sigma N}^{4}\sigma^{4} \\
    &+\frac{1}{2}m_{\rho}^{2}\rho^{2}+3\Lambda_{\nu}g_{\rho N}^{2}g_{\omega N}^{2}\omega^{2}\rho^{2}+ \varepsilon_{K}
\end{aligned}
\end{equation}
where $\varepsilon_{K}$ is the kaonic contribution to the total energy density and is given by:
\begin{equation}
    \varepsilon_{K}= \; \sqrt{{m^*_K}^2 + |q_{K^-}|B} \; \rho_{K}.
\end{equation}
The kaon does not contribute directly to the pressure as it is a (s-wave) Bose condensate so that the expression of pressure reads
\begin{equation}
    P = \;  \mu_p \rho_p + \mu_n \rho_n  + \sum_{l=e, \mu} \mu_l \rho_l  - \varepsilon.
\end{equation}
In addition, it should be stressed that  the electromagnetic contribution to the energy-momentum tensor is cancelled by the Lorentz force associated with magnetization. Thus, although the strong magnetic field induces an anisotropy in the matter part of the energy-momentum tensor, only the isotropic thermodynamic pressure $P$ is relevant for determining equilibrium \cite{Blandford_1982, PhysRevC.85.039801, Chatterjee:2021wsr}.

\begin{table*}
\caption{\label{tab:Table 1}
Parameter sets for the FSUGold model discussed in the text and the meson masses $M_{\sigma}=491.5$ MeV, $M_{\omega}=782.5$ MeV, $M_{\rho}=763$ MeV.
}
\begin{ruledtabular}
\begin{tabular}{lccccccr}
\textrm{Model}&
\textrm{$g_{\sigma}$} &
\textrm{$g_{\omega}$} &
\textrm{$g_{\rho}$} &
\textrm{$\kappa$} &
\textrm{$\lambda$} &
\textrm{$\zeta$} &
\textrm{$\Lambda_{\nu}$}\\
%\hline
FSUGold & 10.59 & 14.30 & 11.77 & 1.42 & 0.0238 & 0.06 & 0.03\\
\end{tabular}
\end{ruledtabular}
\end{table*}

As aforementioned, the problem of the FSUGold model is that the EoS of neutron stars generated by it is not stiff enough, leading to a too small maximum mass of neutron star. In Ref. \cite{Maslov:2015lma}, this problem is solved by the $\sigma$-cut scheme. This scheme points out that, in the range where the density $\rho_B > \rho_0$, a sharp decrease in
the strength of the $\sigma$  meson reduces the decrease in the effective mass of the nucleon, which eventually stiffens the EoS and still yields neutron stars of more than $2M_{Sun}$ after considering the hyperon degrees of freedom \cite{ma2022kaon, ma2023kaon, wu2020hyperonized}.   Recently, the $\sigma$-cut scheme is also successfully implemented to study the properties of high density region and hyperon-rich matter within the relativistic mean-field model using TM1 parameter set \cite{Patra:2022lds}. One of the main goals of this paper is to find out  the effects of $\sigma$-cut scheme on magnetized neutron star matter with kaon condensation.
The $\sigma$-cut scheme  \cite{Maslov:2015lma}, which is able to stiffen the EoS above saturation density, adds in the original Lagrangian density, the function \cite{Maslov:2015lma,Dutra:2015hxa,Kolomeitsev:2015qia}
\begin{eqnarray}
\Delta U(\sigma)=\alpha ln(1+ exp[\beta(f-f_{s,core})] ),
\end{eqnarray}
where $f=g_{\sigma N}\sigma/M_N $ and $f_{s,core}=f_0+c_\sigma (1-f_0)$.  $f_0$ is the value of $f$ at saturation density, equal to 0.61 for the FSUGold model. $\alpha$ and $\beta$ are constants, taken to be $4.822 \times 10^{-4} M_{N}^4$ and 120 as in Ref. \cite{Maslov:2015lma}. The smaller $c_\sigma$ is, the stronger the effect of the $\sigma$-cut scheme becomes. This scheme stiffen the EoS by quenching  the decreasing of the effective mass of nucleon $M_{N}^*=M_{N} (1-f)$ at high density. However, we must be careful so that this scheme do not affect the saturation properties of nuclear matter. In this work, we  refer to the literature \cite{Maslov:2015lma} and adjust the parameter $c_\sigma$  ranging from 0.2 to 0.4  that is able to satisfy the maximum mass constraint.

There is another problem to be considered. It is well known that the TOV equation \cite{oppenheimer1939massive, baldo1997microscopic, kalogera1996maximum} is usually used to get the mass-radius relation of a non-magnetized neutron star. Strictly speaking, the energy-momentum tensor of a neutron star in the presence of strong magnetic field is anisotropic \cite{chatterjee2015consistent, Bocquet:1995je, bonazzola1993axisymmetric, lopes2015magnetized}, and for a central magnetic  field magnitude   stronger than $10^{17}$ G, the spherical symmetry of neutron stars is broken remarkably, then the deformation of neutron stars can be around $2$ percent
 or higher if magnetic field above $10^{17}G$ in the core of neutron stars is used \cite{refId0}, which means in principle the TOV  equation is not reasonable for this situation. Many previous articles simply ignored this issue \cite{rabhi2009quark, menezes2009quark, ryu2010medium, rabhi2011warm, mallick2011possible, lopes2012influence, casali2014hadronic}. A directionally-averaged energy momentum tensor for the magnetic field \cite{lopes2015magnetized} was first proposed in Ref. \cite{bednarek2003influence}. However, the Ref. \cite{chatterjee2019magnetic} suggested this method is less accurate than the simple TOV-like system. Surprisingly, we can get results very close to the exact solution if we just solve the original TOV equation assuming the magnetic field strength is zero. For a central  magnetic field strength $B = 1.0 \times 10^{18}$ G, the maximum mass of neutron stars obtained by this method is only smaller by 0.2\% than the exact solution \cite{chatterjee2019magnetic}. So we will adopt this simple method in our calculation.

Before giving our numerical results, we list parameters for the FSUGold model in Table \ref{tab:Table 1}. The parameter of the models can be found in Ref. \cite{Fattoyev:2010tb,Fattoyev:2010rx,Fattoyev:2010mx} in detail.

\section{Results and discussion}
In this section, we analyze the properties of neutron star matter with kaon condensation in the presence of strong magnetic fields using the FSUGold model. Before the numerical results are shown, we discuss the coupling parameters between  $K^-$ and meson fields. The coupling constants between the vector meson and the Kaon $\emph{g}_{\omega K},g_{\rho K}$ are determined by the meson SU(3) symmetry as $g_{\omega K}=g_{\omega N}/3,g_{\rho K}=g_{\rho N}$ \cite{Char:2014cja}. The scalar coupling constant $g_{\sigma K}$ is fixed to the optical potential of the $\emph{K}^{-}$ in saturated nuclear matter,
\begin{equation}\label{eq3}
\emph{U}_{K}(\rho_{0})=-g_{\sigma K}\sigma(\rho_{0})-g_{\omega K}
\omega(\rho_{0}),
\end{equation}
which characterizes the Kaon-nucleon interaction. Waas and Weise found an attractive potential for the $\emph{K}^{-}$ at the saturation nuclear density of about $\emph{U}_{K}(\rho_{0})=-120$ MeV \cite{Waas:1997pe}.  Another calculation from hybrid model \cite{Friedman:1998xa} suggests the value of $K^{-}$ optical potential to be in the range 180 $\pm$ 20 MeV at saturation density. In this paper, we carry out our calculations with a series of optical potentials ranging from - 160 MeV to -120 MeV \cite{ma2023kaon}.

The $g_{\sigma K}$ can be related to the potential of the kaon at the saturated
density through Eq. (\ref{eq3}). $ g_{\sigma K}$  values corresponding to several values of $\emph{U}_{K}$ are listed in Table \ref{tab:Table 2.}.
\begin{table}
\caption{\label{tab:Table 2.}
$g_{\sigma K}$ determined for several $U_{K}$ values in the FSUGold model.
}
\begin{ruledtabular}
\begin{tabular}{lccr}
\textrm{$U_{K}$ (MeV)} &
-120 &
{-140} &
{-160} \\
%\hline
$g_{\sigma K}$ & 0.768 & 1.358 & 1.948
\end{tabular}
\end{ruledtabular}
\end{table}

As far as  the magnetic field strength is concerned, though the maximum $B$ field in the core of neutron stars might not be higher than  $ 1.0\times 10^{18}$ G from the Virial theorem, we still retain the case of $ 1.0\times 10^{19}$ G for the sake of completeness of the calculation.

\begin{figure}[tbp]
\centering
\includegraphics[width=17cm,height=17cm]{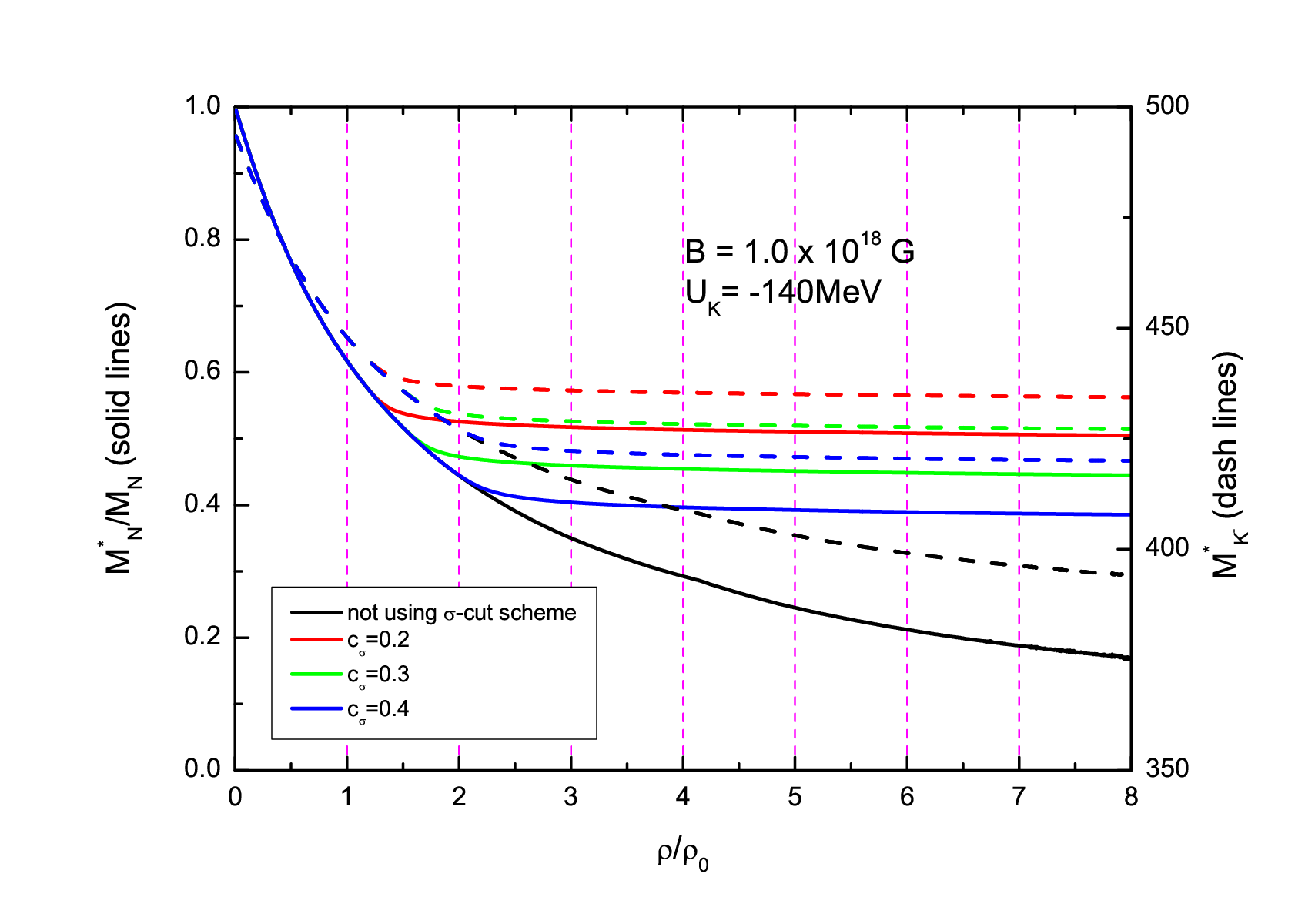}
\caption{effective masses of nucleons and kaons versus baryon density using and not using the $\sigma$-cut scheme for the magnetic field strengths $B = 1.0\times 10^{18}$ G. The solid curves denote  $M_N^*/M_N$, the dashed curves denote $m_K^*$. The values of $m_K^*$ as a function of baryon density are displayed  on the vertical axis to the right of the figure.}
\label{fig:effective_mass}
\end{figure}

First, we want to find out the range for $c_\sigma$ in which the saturation properties of nuclear matter is not affected by the $\sigma$-cut scheme, by examining the effective mass of nucleons under the $\sigma$-cut scheme.
In Figure \ref{fig:effective_mass}, we plot the ratio of the effective mass to the rest mass as a function of baryon density, where $\rho_0$ is the saturation density 0.148 fm$^{-3}$, for the
magnetic field strengths $B = 1.0\times 10^{18}$ G.
We can see that for $c_\sigma \ge 0.2$, the effective mass is unchanged by the $\sigma$-cut scheme when $\rho \le \rho_0$. So the effective mass at saturation density is not affected. Moreover, for $0.2 \le c_\sigma \le 0.4$, the effective mass decreasing is quenched at high baryon density. This is exactly what we want by using the $\sigma$-cut scheme. The smaller $c_\sigma$ is, the stronger the quenching becomes. The effective mass of kaons $m_K^*$ as a function of the baryon density is also displayed  with and without the $c_\sigma$-cut scheme for the magnetic field strengths $B = 1.0\times 10^{18}$ G and $K^-$ potential depth of $U_K$ = -140 MeV.

\begin{figure}[tbp]
\centering
\includegraphics[width=17cm,height=17cm]{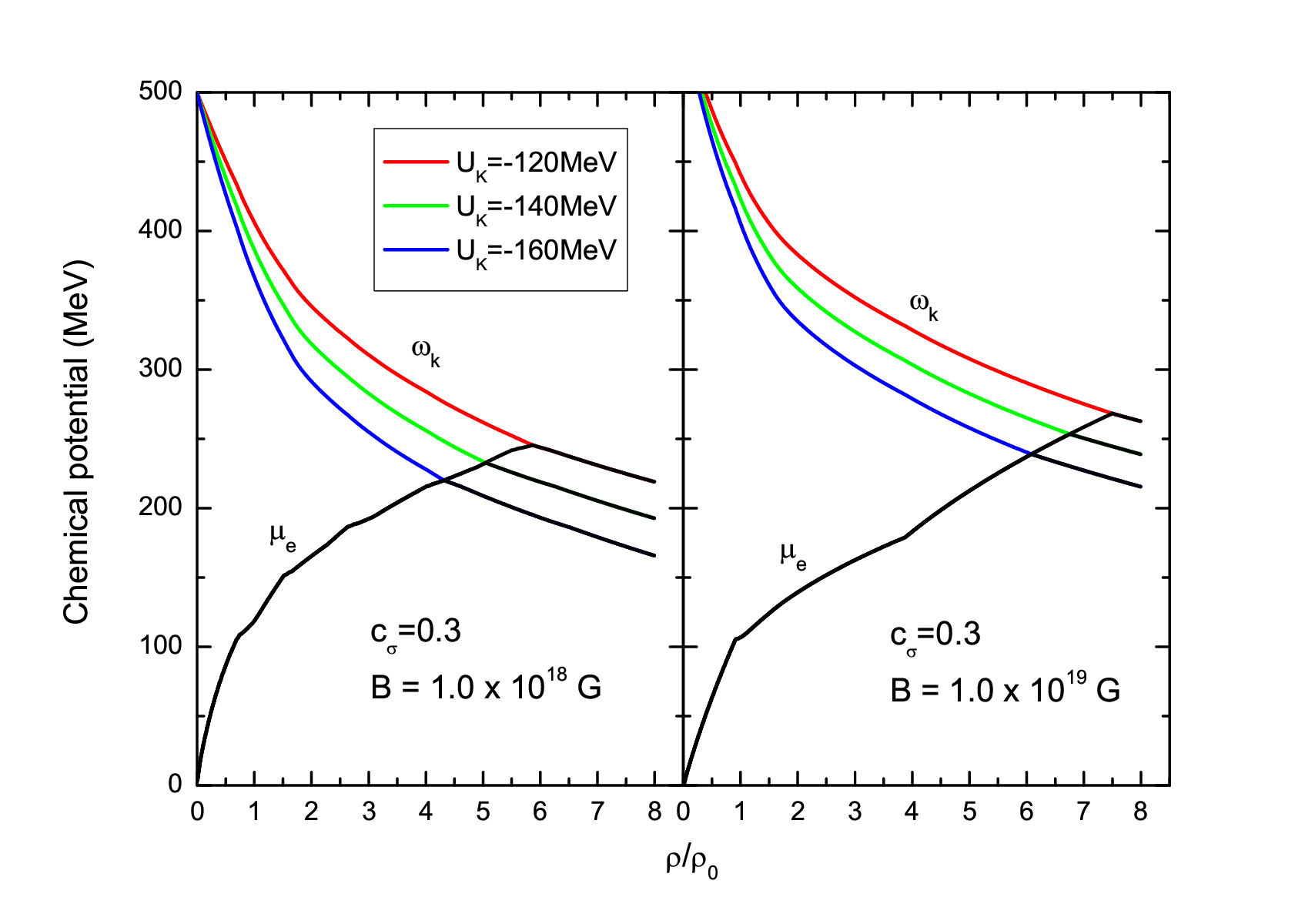}
\caption{Kaon energy ($\omega_K$ ) and electron chemical potential ($\mu_e$) as a function of baryon density for $c_\sigma=0.3$. (Left panel) $B = 1.0 \times 10^{18}$ G; (right panel)  $B = 1.0 \times 10^{19}$ G; The blue curves show $U_K$ = -160 MeV,  green lines show $U_K$ = -140 MeV, and red lines show $U_K$ = -120 MeV.}
\label{fig:chem}
\end{figure}

Figure \ref{fig:chem} shows the Kaon energy ($\omega_{k}$) and electron chemical potential ($\mu_{e}$) as a function of baryon density with $U_{K} = -120, -140, -160$ MeV for the parameter $B = 1.0 \times 10^{18}$ G and  $ 1.0 \times 10^{19}$ G. $K^{-}$ condensation initiates once the value of $\omega_{K}$ reaches that of the electron chemical potential.

\begin{figure}[tbp]
\centering
\includegraphics[width=17cm,height=17cm]{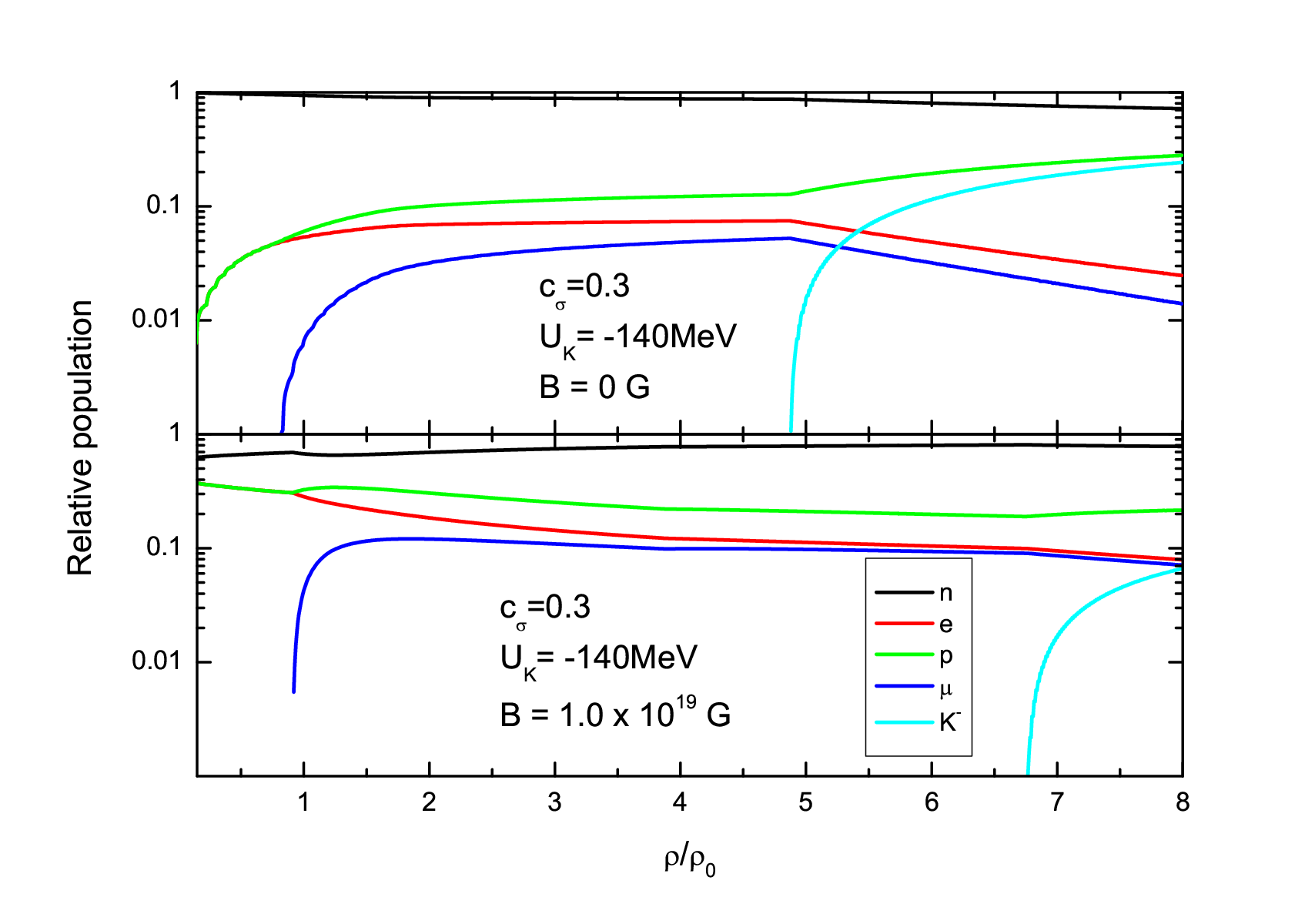}
\caption{Relative population of particles versus baryon density for $K^-$ optical potential depth of $U_K = -$ 140 MeV and $c_\sigma=0.3$. (Upper panel) $B =0$ G; (lower panel) $B = 1.0 \times 10^{19}$ G.}
\label{fig:population}
\end{figure}

Figure \ref{fig:population} shows the relative population of particles versus baryon density with Kaon optical potential $U_{K}= -140$ MeV and  $c_\sigma =0.3$. For $B = 0$ G, the mixed phase initiates with the onset of $K^{-}$ at $\sim 4.9 \rho_{0}$; for $B = 1.0\times 10^{19}$ G, with the appearance of $K^{-}$ at $\sim 6.8 \rho_{0}$. Here we note that the formation of $K^-$ condensation in the presence of strong field is delayed to higher density than the field-free case. This is mainly because the negatively charged $K^-$ gets a large chemical potential in the presence of strong magnetic field, as given in Eq. (\ref{kaon chemical}) due to the term $|q_{K^-} |B$.

\begin{figure}[tbp]
\centering
\includegraphics[width=17cm,height=17cm]{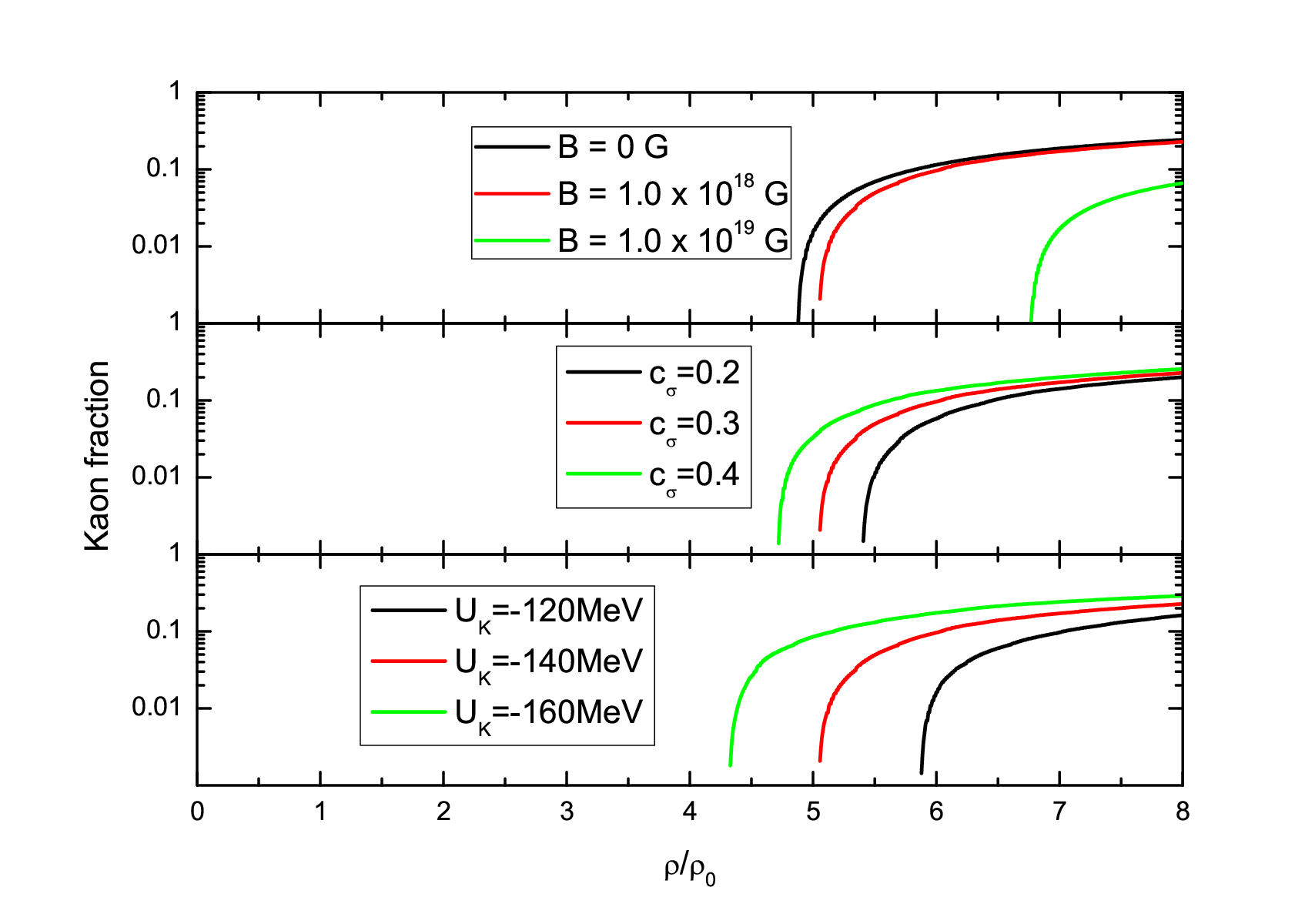}
\caption{The kaon fraction as a function of the baryon density. (Upper panel) for $B =0, 1.0 \times 10^{18}$, and $1.0 \times 10^{19}$ G  with fixed $U_K = -$ 140 MeV and $c_\sigma=0.3$; (middle panel) for $c_\sigma= 0.2, 0.3$ and 0.4  with fixed $U_K = -$ 140 MeV and $B = 1.0 \times 10^{18}$ G; (lower panel) for $U_K = -$ 120, $-$140 and $-$160 MeV   with fixed $B = 1.0 \times 10^{18}$ G and $c_\sigma=0.3$.}
\label{fig:Kaon_fraction}
\end{figure}

In figure \ref{fig:Kaon_fraction}, we present the $K^-$ fraction as  a function of the baryon density for different parameters. From the middle panel of this figure, we can find that the percentage of $K^-$ is decreased by
the $\sigma$-cut scheme. We list the threshold densities $\rho_{cr}$ for kaon condensation for different values of $K^-$ optical potential depth $U_K$, the $\sigma$-cut scheme parameter $c_\sigma$ and magnetic field strength $B$ in
Table \ref{tab:threshold UK.}, Table \ref{tab:threshold sigma cut.} and Table \ref{tab:threshold B field.}.

\begin{table}
\caption{\label{tab:threshold UK.}
Threshold densities $\rho_{cr}$ (in units of the nuclear saturation density $\rho_0$) for kaon condensation in dense nuclear matter for different values of $K^-$ optical potential depths $U_K$ (in units of MeV) with $c_\sigma = 0.3$ and $B = 1.0 \times 10^{18}$ G.
}
\begin{ruledtabular}
\begin{tabular}{lccr}
\textrm{$U_{K}$ (MeV)} &
-120 &
{-140} &
{-160} \\
%\hline
$\rho_{cr}$($K^-$) & 5.880 & 5.056 & 4.328
\end{tabular}
\end{ruledtabular}
\end{table}

\begin{table}
\caption{\label{tab:threshold sigma cut.}
Threshold densities $\rho_{cr}$ (in units of the nuclear saturation density $\rho_0$) for kaon condensation in dense nuclear matter for different values of   $c_\sigma $ with $U_K = - 140$ MeV and $B = 1.0 \times 10^{18}$ G.}
\begin{ruledtabular}
\begin{tabular}{lcccr}
\textrm{$c_\sigma$ } &
without the $\sigma$-cut scheme &
0.2 & 0.3 &0.4 \\
$\rho_{cr}$($K^-$) & 4.144 & 5.408 & 5.056 &4.720
\end{tabular}
\end{ruledtabular}
\end{table}

\begin{table}
\caption{\label{tab:threshold B field.}
Threshold densities $\rho_{cr}$ (in units of the nuclear saturation density $\rho_0$) for kaon condensation in dense nuclear matter for different values of magnetic field strength  $B$  with $c_\sigma = 0.3$ and $U_K = - 140$ MeV.
}
\begin{ruledtabular}
\begin{tabular}{lccr}
B (G)  & 0 & $ 1.0 \times 10^{18}$ & $ 1.0 \times 10^{19}$ \\
%\hline
$\rho_{cr}$($K^-$) & 4.856 & 5.056 & 6.768
\end{tabular}
\end{ruledtabular}
\end{table}

\begin{figure}[tbp]
\centering
\includegraphics[width=17cm,height=17cm]{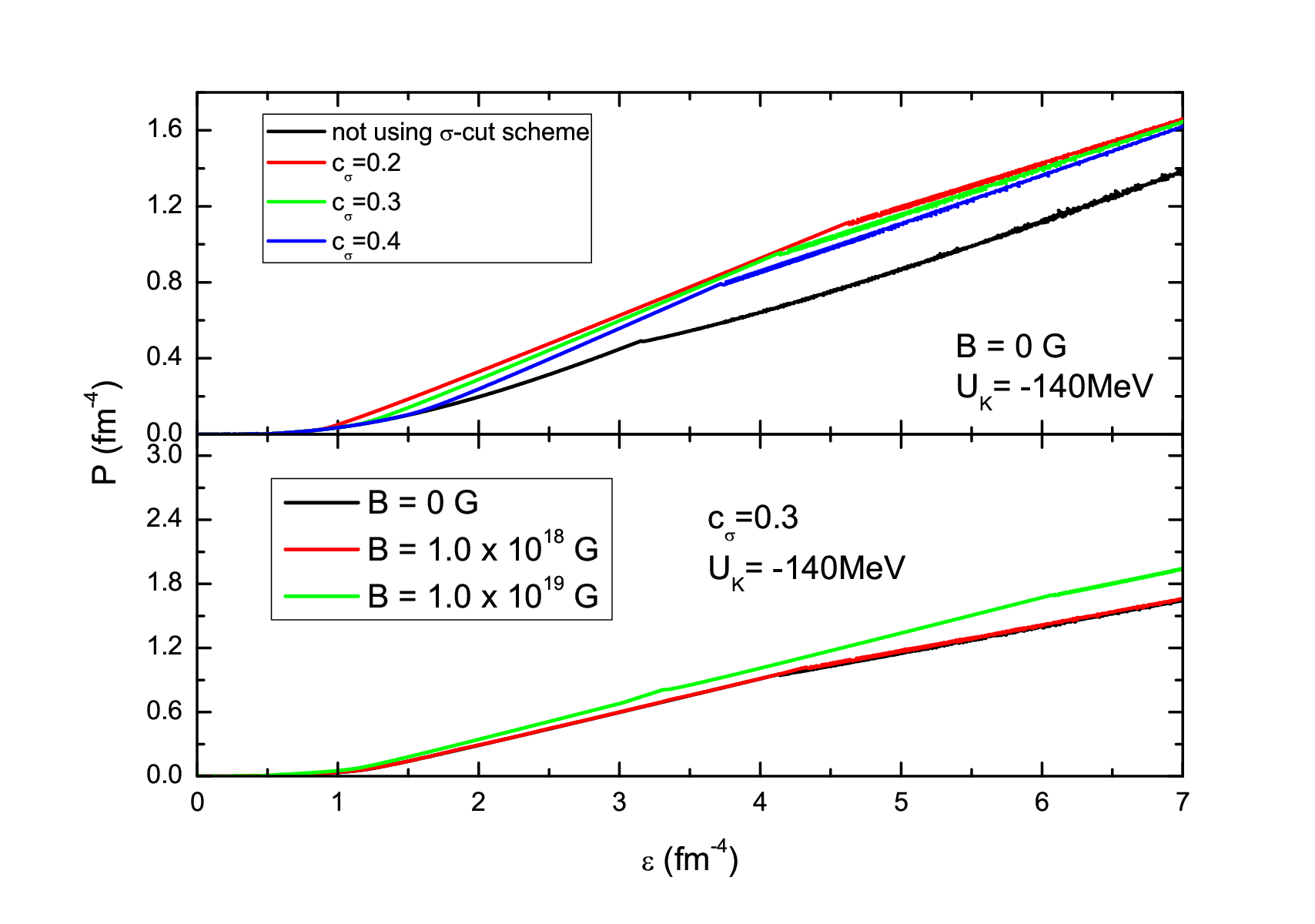}
\caption{Pressure as a function of energy density. (Upper panel) using and not using the $\sigma$-cut scheme with fixed  $U_K = -$140 MeV  and $B =0$ G; (lower panel) for $B = 0, 1.0 \times 10^{18}$, and $1.0 \times 10^{19}$ G with fixed  $U_K = -$140 MeV  and $c_\sigma = 0.3$.}
\label{fig:eos}
\end{figure}

The $\sigma$-cut scheme works by stiffening the EoS of neutron star matter. We compare the EoS of neutron star matter between
the case that does not use the $\sigma$-cut scheme and the case that uses it in  upper panel of Figure \ref{fig:eos}. Indeed, the EoS is stiffened by the $\sigma$-cut scheme significantly.
The EoS is stiffer for smaller $c_\sigma$ as expected. In  lower panel of Figure  \ref{fig:eos}, we show the matter pressure $P$ as a function of the matter energy density $\epsilon$ for the magnetic field strengths $B = 0, 1.0 \times 10^{18}$, and $1.0 \times 10^{19}$ G. At higher nuclear densities, the influence of strong magnetic fields on the EoS becomes noticeable. The threshold of kaon condensation shifts to higher density and the effect of kaon
condensation on the EoS gets weaker with increasing $B$. Moreover, the EoSs for the magnetic field strengths $B = 0, 1.0 \times 10^{17}, 1.0 \times 10^{18}$, and $1.0 \times 10^{19}$ G are plotted for original FSU model without the $\sigma$-cut scheme. As one can see,  the EoSs  in the presence of weak magnetic fields will only have a negligible difference relative to the field-free case, which is qualitatively consistent with other calculations \cite{Chatterjee:2021wsr, PhysRevC.110.045805}.

\begin{figure}[tbp]
\centering
\includegraphics[width=17cm,height=17cm]{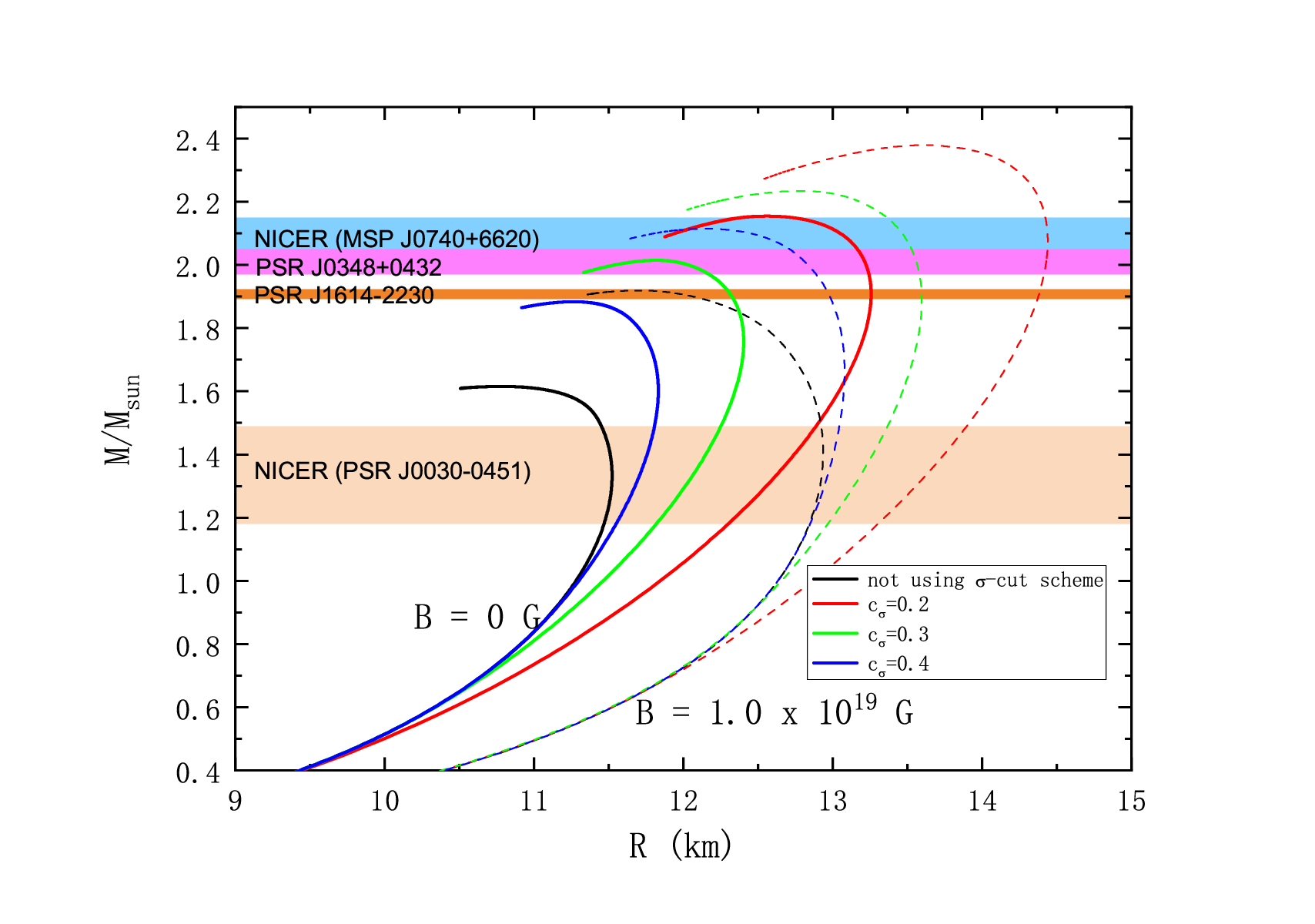}
\caption{Mass-radius relation using and not using $\sigma$-cut scheme in neutron star matter including kaon condensation. The solid lines denote $B = 0$ G, dashed lines denote $B= 1.0 \times 10^{19}$ G, and $U_{K}=-140$ MeV. The horizontal bars indicate the observational constraints of PSR J1614 - 2230 \cite{Demorest:2010bx,ozel2010massive}, PSR J0348 + 0432 \cite{antoniadis2013j}, MSP J0740 + 6620 \cite{fonseca2021refined,cromartie2020relativistic} and PSR J0030-0451 \cite{Riley_2019}.}
\label{fig:mass-radius}
\end{figure}

Next the results of mass-radius relation for static spherical stars from solution of the TOV equation discussed here are shown in Fig. \ref{fig:mass-radius}. The mass measurements of PSR J1614 - 2230 \cite{Demorest:2010bx,arzoumanian2018nanograv,fonseca2016nanograv,ozel2010massive}, PSR J0348 + 0432 \cite{antoniadis2013j}, MSP J0740 + 6620 and PSR J0030 - 0451 are indicated by the horizontal bars. We find that the $\sigma$-cut scheme can significantly increase the maximum mass of the neutron star, the smaller $c_{\sigma}$ is, the stronger the effect of this scheme is. The magnetic field strength $B$ will  significantly affect the maximum mass of the neutron star when $B= 1.0 \times 10^{19}$ G.

\begin{figure}[tbp]
\centering
\includegraphics[width=17cm,height=17cm]{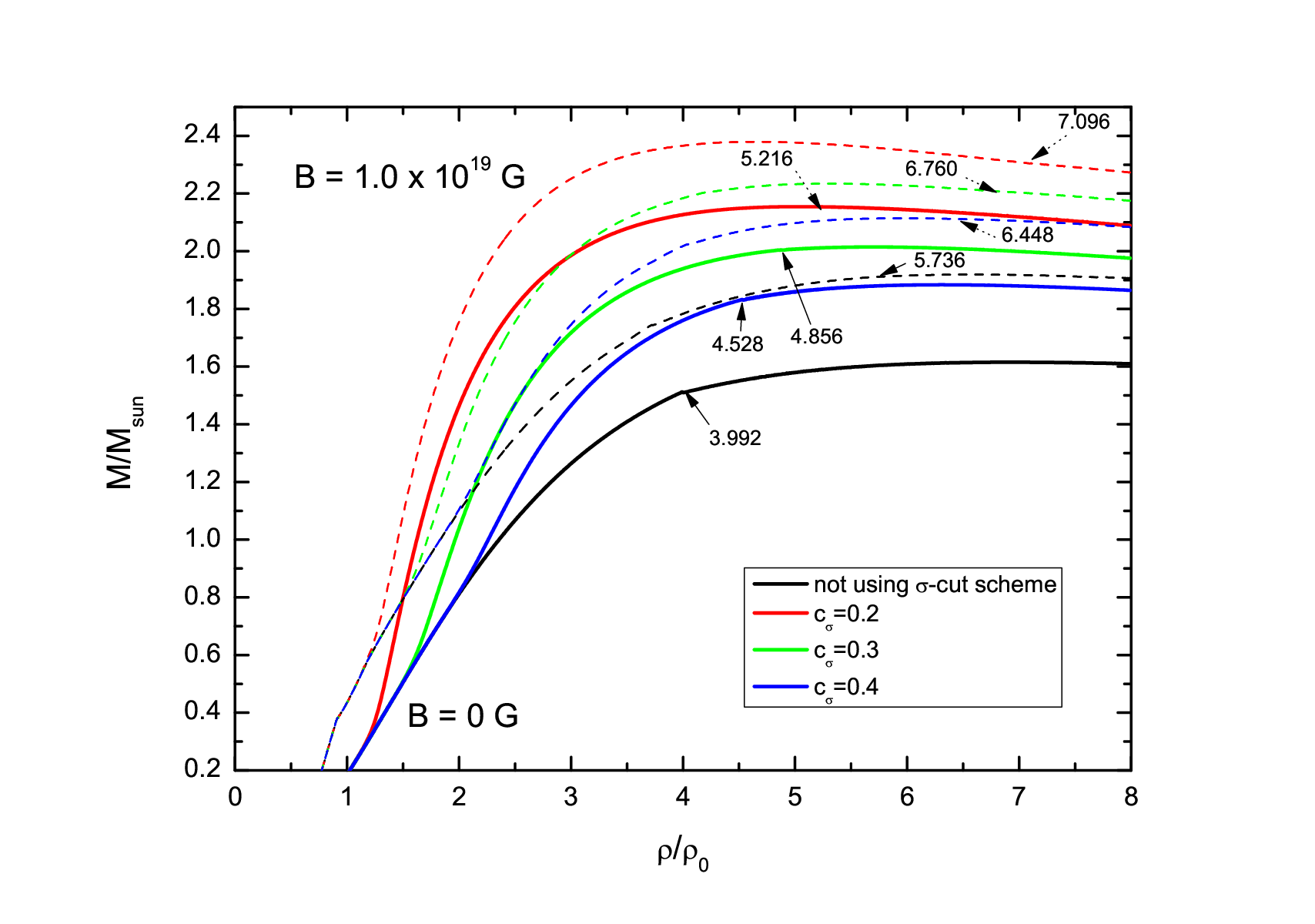}
\caption{Mass-central density relation using and not using $\sigma$-cut scheme in neutron star matter including kaon condensation. The solid lines denote $B = 0$ G, dashed lines denote $B= 1.0 \times 10^{19}$ G, and $U_{K}=-140$ MeV. Arrow point to the onset density of kaons. Dotted arrow indicate that the onset density exceeds the central density of the maximum mass star, so kaon condensation does not occur in such case.}
\label{fig:m(rho)}
\end{figure}

The existence of kaon condensation depends crucially on the relationship between the critical density and the central density of the maximum mass stars. Fig. \ref{fig:m(rho)} shows Mass-central density relation using and not using $\sigma$-cut scheme in neutron star matter. The solid lines denote $B = 0$ G, dashed lines denote $B= 1.0 \times 10^{19}$ G, and $U_{K}=-140$ MeV. The corresponding critical density of the kaon condensate is also shown in this figure. Arrow point to the onset density of kaons in neutron star matter. Dotted arrow indicate that the onset density exceeds the central density of the maximum mass star, so kaon condensation does not occur in such case.
  As one can see, the critical density is shifted right significantly in the presence of magnetic field.  For the $B = 1.0 \times 10^{19}$ case, only when the sigma-cut scheme is not used, can kaon condensation appear.
These interesting phenomena have also been discussed in the literature \cite{Sedrakian:2022ata}.
\section{Summary}

In this paper, the FSUGold model with the inclusion of kaon condensation in the presence of strong magnetic field is used to study effects of the $\sigma$-cut scheme. The original model generate an EoS too soft to produce a maximum mass neutron star heavier than 2.01 $M_{sun}$. Applying the $\sigma$-cut scheme, with careful choice of the parameter of this scheme, we got the maximum mass heavier than $2M_{sun}$ within the range of observed mass measurements.
 We have studied the effects of strong magnetic field on kaon condensation in neutron star matter. We found that the presence of strong magnetic field significantly alters the threshold density of kaon condensation. The threshold of kaon condensation shifts to higher density in the presence of strong magnetic field. For $B = 0, 1.0 \times 10^{18}$, and $1.0 \times 10^{19}$ G with fixed $U_K = -$ 140 MeV and $c_\sigma=0.3$, we obtained the threshold densities of $K^-$ condensation are 4.856, 5.056  and 6.768 (in units of the nuclear saturation density $\rho_0$). It is obvious that the threshold density of $K^-$ condensation depends strongly on the magnetic field strength.

In general, a neutron star consists of an inner crust of nuclei in a gas of neutrons and electrons, an outer crust of nuclei in a gas of electrons, and a liquid core of uniform dense matter. The inner crust of neutron stars has drawn much attention due to its complex phase structure and significant role in astrophysical observations \cite{PhysRevC.94.035804}. Taking the physics of crust properties into consideration is a issue  of great academic significance. We hope to investigate the physics of crust properties within the framework of this model in the near future.

\begin{acknowledgments}
This work was supported by the Zhejiang normal university Doctoral research fund Contract No. ZC302924005.
\end{acknowledgments}

\bibliography{ns}

\begin{figure}[tbp]
\centering
\includegraphics[width=17cm,height=17cm]{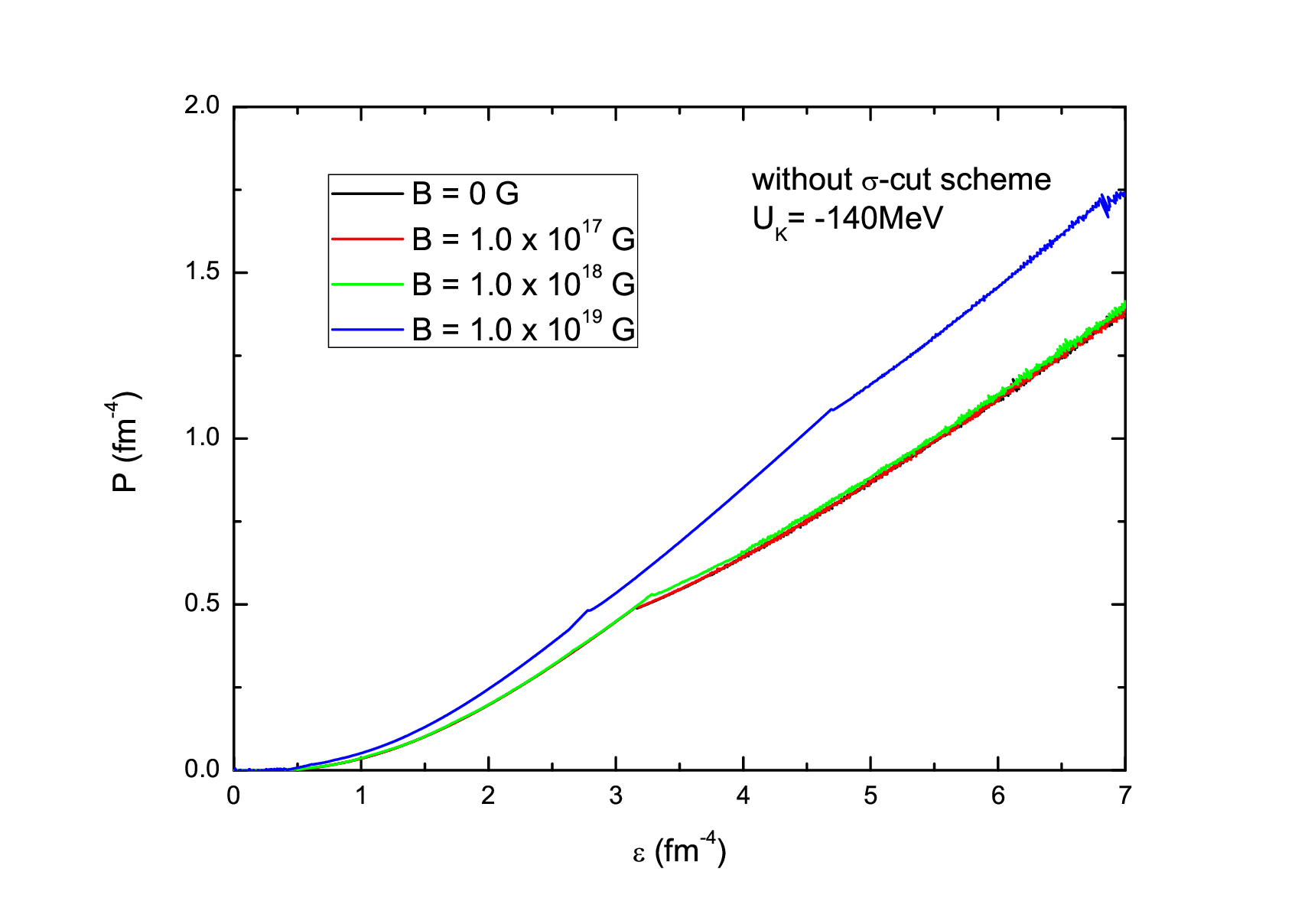}
\caption{Pressure as a function of energy density without the $\sigma$-cut scheme for different values of $B$ with fixed  $U_K = -$140 MeV.}
\label{fig:eos_no_cut}
\end{figure}

\end{document}